\documentclass[runningheads]{llncs}
\usepackage[T1]{fontenc}
\usepackage{graphicx}
\graphicspath{{figures/}}
\usepackage{hyperref}
\IfFileExists{xurl.sty}{\usepackage{xurl}}{}
\usepackage{longtable}
\usepackage{array}
\usepackage{underscore}
\usepackage{listings}
\usepackage{xcolor}
\usepackage{marvosym}
\let\Envelope\Letter
\usepackage{booktabs}

\hypersetup{
  colorlinks=true,
  linkcolor=blue,
  citecolor=blue,
  urlcolor=blue
}

\definecolor{MyDarkBlue}{RGB}{10, 10, 185}
\definecolor{MyDarkRed}{RGB}{175, 0, 0}
\definecolor{MyDarkGreen}{RGB}{0, 175, 60}
\definecolor{MyCyan}{RGB}{20, 145, 145}
\definecolor{MyBrown}{RGB}{165,42,42}
\newcommand{\CodeSymbolGreen}[1]{\textcolor{MyDarkGreen}{#1}}

\newcommand{\CodeSymbolCyan}[1]{\textcolor{MyCyan}{#1}}
\newcommand{\CodeSymbolRed}[1]{\textcolor{MyDarkRed}{#1}}
\newcommand{\CodeSymbolBrown}[1]{\textcolor{MyBrown}{#1}}

\lstdefinestyle{sparqlStyle}
{   basicstyle=\tiny\ttfamily,
    breaklines=true,
    numbers=none,
    keywordstyle=\color{MyDarkBlue}\bfseries,
    keywords=[1]{@prefix,PREFIX,@base,BASE,SELECT,DISTINCT,ORDER,BY,VALUES,FILTER,WHERE,UNION,GROUP,BIND,INSERT,DELETE,ASK,SERVICE},
    morekeywords=[2]{{[},{]}},
    morekeywords=[2]{{\{},{\}}},
    comment=[l]{\#},
    morecomment=[s][\color{MyDarkBlue}]{<}{>},
    tabsize=4,
    alsoletter={-?},
    numberstyle=\color{black},
    morestring=[b][\color{black}]",
    showstringspaces=false,
    literate= {;}{{\CodeSymbolRed{;}}}1
     {.}{{\CodeSymbolRed{.}}}1
     {"}{{\CodeSymbolRed{"}}}1
     {\{}{{\CodeSymbolGreen{\{}}}1
     {\}}{{\CodeSymbolGreen{\}}}}1
     {]}{{\CodeSymbolCyan{]}}}1
     {[}{{\CodeSymbolCyan{[}}}1
     {(}{{\CodeSymbolBrown{(}}}1
     {)}{{\CodeSymbolBrown{)}}}1,
    moredelim=[s][\color{MyDarkBlue}]{:}{\ },
    moredelim=[s][\color{MyDarkRed}]{@}{\ },
}

\begin{document}
\title{MetaboKG: An Analysis-centric Knowledge Graph Framework for Untargeted Metabolomics}
\titlerunning{MetaboKG}
\author{Matthieu F\'eraud\inst{1,2}\orcidID{0009-0001-5496-6690}\Envelope \and
Dina Boukhajou\inst{1,2}\orcidID{0009-0006-2243-7010} \and
Fabien Gandon\inst{2,3}\orcidID{0000-0003-0543-1232}\Envelope \and
Louis-F\'elix Nothias\inst{1,2}\orcidID{0000-0001-6711-6719}\Envelope}
\authorrunning{M. F\'eraud et al.}
\institute{Univ.\ C\^ote d'Azur, CNRS, ICN, France\\
\email{matthieu.feraud@univ-cotedazur.fr, louis-felix.nothias@cnrs.fr}
\and
Interdisciplinary Institute for Artificial Intelligence (3iA) C\^ote d'Azur, Sophia-Antipolis, France
\and
Inria, Univ.\ C\^ote d'Azur, CNRS, I3S, France\\
\email{fabien.gandon@inria.fr}}

\maketitle              
\begin{abstract}
Untargeted metabolomics generates large volumes of tandem mass spectrometry (MS/MS) data and computational annotations that can reveal molecular mechanisms across organisms and environments. Public reuse has improved through harmonized repository metadata and access infrastructures such as Pan-ReDU, and through metabolomics knowledge graphs such as ENPKG and METRIN-KG. Yet the analytical layer remains fragmented: spectra, features, workflow outputs, annotations, confidence evidence, and contextual metadata are still scattered across repositories and tabular artifacts. We present MetaboKG, an analysis-centric knowledge graph framework for engineering reusable metabolomics knowledge from public repositories, metadata, and GNPS molecular network results. MetaboKG contributes a transformation workflow that preserves links between repository exports, analytical files, spectra, features, and annotation results; a semantic model grounded in PROV-O and SIO and aligned with the Mass Spectrometry ontology (MS), ChEBI, NCBITaxon, ENVO, and NCIT to represent provenance, analytical evidence, metadata attributes, and controlled vocabulary terms; and a Universal Annotation Identifier strategy extending the Universal Spectrum Identifier (USI) with workflow-specific components for late binding, incremental ingestion, and post hoc linkage across analyses. We demonstrate MetaboKG at the public-repository scale on 680 GNPS molecular networking results and evaluate it through competency questions covering biochemical enrichment, environmental specificity, and cross-instrument analytical variation. Results show that graph-based integration supports traceable annotation reuse and reproducible SPARQL exploration of biochemical relationships that remain fragmented across repository-native resources.
\end{abstract}

\section{Introduction}

Untargeted metabolomics is a major instrument of contemporary chemical and life-science research, with applications ranging from natural product discovery to functional genomics and phenotype characterization~\cite{atanasovNaturalProductsDrug2021,fiehnMetabolomicstheLinkGenotypes2002,commissoUntargetedMetabolomicsEmerging2013}. It relies on tandem mass spectrometry (MS/MS) to detect molecular signatures in complex samples~\cite{boccardKnowledgeDiscoveryMetabolomics2010,manierUntargetedMetabolomicsHigh2019}, and as public resources such as GNPS/MassIVE~\cite{wangSharingCommunityCuration2016}, MetaboLights~\cite{haugMetaboLightsResourceEvolving2020}, and the Metabolomics Workbench continue to grow~\cite{sudMetabolomicsWorkbenchInternational2016}, the field is moving from experiment-centered processing toward cross-study reuse and large-scale computational analysis~\cite{santamariaBioinformaticAnalysisMetabolomic2024,elabieadEnablingPanrepositoryReanalysis2025,chenKnowledgeGraphsLife2023}. These opportunities depend on the ability to combine heterogeneous sources while preserving experimental context, provenance, and interpretation~\cite{wilkinsonFAIRGuidingPrinciples2016,cambiaghiAnalysisMetabolomicData2017,kleinsteuberManagingProvenanceData2024}.

The main difficulty is that untargeted metabolomics data remain fragmented across repositories that follow different submission logics, file formats, and metadata practices~\cite{haugMetaboLightsResourceEvolving2020,wangSharingCommunityCuration2016}. GNPS/MassIVE emphasizes community-scale sharing and, through ReDU and Pan-ReDU, has made significant progress toward harmonized metadata~\cite{jarmuschReDUFrameworkFind2020,elabieadEnablingPanrepositoryReanalysis2025}, though annotation reporting remains uneven across submissions, while MetaboLights and the Metabolomics Workbench rely on more structured metadata yet often omit analytical details crucial for interpreting annotations and confidence scores~\cite{yurektenMetaboLightsOpenData2024,sudMetabolomicsWorkbenchInternational2016,metzIntroducingIdentificationProbability2025}. Researchers can therefore inspect results within a single workflow, but the table-based representation of annotation outputs amplifies this fragmentation: heterogeneous column structures and implicit conventions make it difficult to align datasets, compare analyses, or reuse annotations across platforms~\cite{cambiaghiAnalysisMetabolomicData2017,stancliffeQuickTipsReusing2022,elabieadEnablingPanrepositoryReanalysis2025}. The limited interoperability of table-based representations and the underuse of biological and environmental context further amplify this fragmentation~\cite{hoffmannMzTabMDataStandard2019,walkerFunctionalTraits202022,deloryTraitbasedFrameworkLinking2024}.

This raises a general question: \textit{do current semantic web languages and ontologies cover the needs to integrate semi-structured MS/MS data from heterogeneous repositories as knowledge graphs?} We refine it into three sub-questions:
\begin{itemize}
    \item \textbf{RQ1.} how can semi-structured MS/MS data, repository exports, and heterogeneous metadata be transformed into a coherent graph structure?
    \item \textbf{RQ2.} how can Semantic Web technologies and selected ontologies be used to harmonize metadata, preserve provenance, and improve FAIR compliance?
    \item \textbf{RQ3.} how can such a graph support inference and cross-dataset exploration to reveal biochemical relationships that remain hidden when resources are analyzed in isolation?
\end{itemize}

We address these questions through MetaboKG, an analysis-centric knowledge graph framework that connects analytical outputs with contextual metadata while keeping provenance explicit. Section~\ref{sec:related-work} positions our contribution. Section~\ref{sec:modeling} describes the modeling strategy, Section~\ref{sec:collecting} presents the transformation workflow over Pan-ReDU and GNPS results, Section~\ref{sec:provenance} discusses provenance and FAIR compliance, and Section~\ref{sec:evaluation} evaluates the framework through competency questions. Section~\ref{sec:conclusion} concludes.

\section{Related Work}
\label{sec:related-work}
The design of MetaboKG sits at the intersection of three trends: metabolomics data standardization, large-scale reuse of public metabolomics repositories, and knowledge-graph-based integration in the life sciences~\cite{hoffmannMzTabMDataStandard2019,elabieadEnablingPanrepositoryReanalysis2025,chenKnowledgeGraphsLife2023}. No existing approach jointly addresses heterogeneous untargeted metabolomics results, analytical provenance, and cross-repository exploration~\cite{nothiasFeaturebasedMolecularNetworking2020,duhrkopSIRIUS4Rapid2019,santosKnowledgeGraphInterpret2022}.

\paragraph{From metabolomics exchange formats to cross-repository reuse.}
Raw mass spectrometry data are now relatively well supported by established formats~\cite{martensMzMLCommunityStandard2011,hoffmannMzTabMDataStandard2019}, but downstream annotation results and sample-level metadata remain heterogeneous because GNPS/MassIVE, MetaboLights, and Metabolomics Workbench expose complementary but uneven descriptions of studies, samples, and annotations~\cite{wangSharingCommunityCuration2016,haugMetaboLightsResourceEvolving2020,sudMetabolomicsWorkbenchInternational2016,santamariaBioinformaticAnalysisMetabolomic2024}. ISA-Tab~\cite{johnsonISAAPIOpen2021}, the Metabolomics Standards Initiative~\cite{sansoneMetabolomicsStandardsInitiative2007,goodacreProposedMinimumReporting2007}, and ReDU/Pan-ReDU~\cite{jarmuschReDUFrameworkFind2020,elabieadEnablingPanrepositoryReanalysis2025} create a first layer of comparability across datasets, but harmonized tables remain difficult to connect to heterogeneous analysis outputs in a reusable way and stay centered on data delivery rather than integrated knowledge exploration~\cite{cambiaghiAnalysisMetabolomicData2017,chenKnowledgeGraphsLife2023}. mzTab-M illustrates this limit: it aimed to unify metabolomics result representation, but its tabular structure is not well suited to linking diverse annotation outputs, provenance, and contextual metadata in a single queryable model~\cite{hoffmannMzTabMDataStandard2019,metzIntroducingIdentificationProbability2025,duhrkopSystematicClassificationUnknown2021}.

\sloppy
\paragraph{Evolution of annotation workflows in untargeted metabolomics.}
Annotation has evolved from spectral-library matching toward a more diverse ecosystem of workflows~\cite{chaleckisChallengesProgressPromises2019,xuUnveilingDarkMatter2025}. GNPS and feature-based molecular networking structured community-scale annotation sharing~\cite{wangSharingCommunityCuration2016,nothiasFeaturebasedMolecularNetworking2020}, while SIRIUS, CANOPUS, and CSI:FingerID democratized machine-learning-assisted annotation through probabilistic structure and class prediction~\cite{duhrkopSearchingMolecularStructure2015,duhrkopSIRIUS4Rapid2019,duhrkopSystematicClassificationUnknown2021}. These workflows produce multiple candidate structures, class predictions, and software-specific intermediate results, so the integration problem is no longer only metadata harmonization but also the joint representation of heterogeneous analytical claims with the context needed to interpret them~\cite{vaniyaUsingFragmentationTrees2015,metzIntroducingIdentificationProbability2025,bittremieuxCriticalRoleThat2022,chenKnowledgeGraphsLife2023,santosKnowledgeGraphInterpret2022}. A graph-based representation makes it possible to preserve parallel annotations, attach them to their generating activities, and connect them to samples, datasets, and controlled vocabulary terms~\cite{chenKnowledgeGraphsLife2023,galgonekIDSMMassSpectrometry2024,kleinsteuberManagingProvenanceData2024}.

\paragraph{Knowledge graphs in the life sciences and in metabolomics.}
Knowledge graphs have already demonstrated their value for integrating heterogeneous and interdependent data in the life sciences, notably in genomics, proteomics, drug discovery, and precision medicine~\cite{chenKnowledgeGraphsLife2023,ashburnerGeneOntologyTool2000,santosKnowledgeGraphInterpret2022}. Their main strength is to provide a flexible semantic layer in which entities from multiple sources can be linked while retaining explicit meanings and relations~\cite{chenKnowledgeGraphsLife2023,dumontierSemanticscienceIntegratedOntology2014}. This makes them particularly relevant for domains where data are fragmented across formats, repositories, and levels of interpretation~\cite{chenKnowledgeGraphsLife2023,cambiaghiAnalysisMetabolomicData2017}.
In metabolomics and natural product research, several recent projects show the growing importance of this paradigm. LOTUS demonstrated the value of graph-based aggregation for structure--organism knowledge at large scale~\cite{rutzLOTUSInitiativeOpen2022a}. ENPKG then moved closer to experimental metabolomics by integrating annotation results, taxonomic information, and chemical knowledge for natural product research~\cite{gaudrySampleCentricKnowledgeDrivenComputational2024}. METRIN-KG further extended this perspective by connecting metabolomic information with trait and interaction data, illustrating the broader trend toward ecologically and biologically contextualized metabolomics knowledge graphs~\cite{tandonMETRINKGKnowledgeGraph2026}.

These contributions show that semantic technologies can support richer exploration of metabolomics data than repository-native tables alone~\cite{gaudrySampleCentricKnowledgeDrivenComputational2024,tandonMETRINKGKnowledgeGraph2026}. But they leave an open space that MetaboKG addresses~\cite{chenKnowledgeGraphsLife2023}. Existing metabolomics-oriented knowledge graphs generally focus on curated collections, specific experimental programs, or downstream biological interpretation layers~\cite{rutzLOTUSInitiativeOpen2022a,gaudrySampleCentricKnowledgeDrivenComputational2024,tandonMETRINKGKnowledgeGraph2026}. They do not primarily target the systematic integration of heterogeneous public untargeted metabolomics repositories together with their repository-native analysis outputs and metadata heterogeneity~\cite{wangSharingCommunityCuration2016,yurektenMetaboLightsOpenData2024,sudMetabolomicsWorkbenchInternational2016}. In addition, provenance is often present only partially or is tailored to a specific workflow, whereas our objective is to make analytical origin, file-level traceability, and cross-workflow linkage central design principles~\cite{kleinsteuberManagingProvenanceData2024,galgonekIDSMMassSpectrometry2024}.

\paragraph{Positioning of MetaboKG.}
Taken together, the literature suggests that an analysis-centric framework is still needed to connect public untargeted metabolomics repositories, heterogeneous annotation outputs, and contextual metadata in a unified provenance-aware representation~\cite{kleinsteuberManagingProvenanceData2024,elabieadEnablingPanrepositoryReanalysis2025}.

As a response, MetaboKG  builds on prior standardization efforts rather than replacing them, reuses harmonized metadata resources such as Pan-ReDU, and follows the broader trend toward knowledge-graph-based metabolomics integration~\cite{hoffmannMzTabMDataStandard2019,elabieadEnablingPanrepositoryReanalysis2025,chenKnowledgeGraphsLife2023}. Its specific contribution is to shift the focus from isolated repository exports or workflow-specific tables toward a common graph structure where analytical entities, metadata, and provenance can be queried jointly~\cite{kleinsteuberManagingProvenanceData2024,galgonekIDSMMassSpectrometry2024}.

\section{Modeling a Knowledge Graph for Untargeted Metabolomics}
\label{sec:modeling}

To maintain conceptual consistency and reduce ontology alignment overhead, MetaboKG deliberately relies on a limited set of complementary ontologies with clear hierarchical roles. The rest of this section motivates this stack and the modeling patterns that hold it together.

MetaboKG is grounded in the PROV-O ontology, which represents provenance entities and documents how data were collected, enabling meaningful reuse \cite{PROVOPROVOntology}. Although we initially considered using SOSA, given its emphasis on sensing and sampling activities, we ultimately did not adopt it \cite{SemanticSensorNetwork}. While PROV-O and SOSA are not inherently incompatible, MetaboKG is intentionally designed to prioritize representation of the data workflow rather than biological transformation processes. SOSA would have forced us to describe in detail biological processes such as sample collection or extraction, which would have bloated the knowledge graph without yielding any practical benefits.

To structure relations between entities and their qualifying values, we adopt the n-ary relation pattern by introducing an intermediate resource between the subject and the value of a property \cite{noyDefiningNaryRelations2006}. Rather than committing to a context-specific datatype property for each value, the intermediate resource carries the value together with its semantic context through its \texttt{rdf:type}, which can be drawn from standard domain ontologies (MS, CHMO, NCIT, ChEBI, etc.). The predicates linking subjects to these intermediate resources are taken from the SIO ontology, which provides generic relations together with their inverse counterparts \cite{dumontierSemanticscienceIntegratedOntology2014}. Reusing SIO predicates reduces maintenance overhead, since their inverse relations are already declared.

To limit graph size and avoid unnecessary duplication of intermediate resources, we apply two complementary strategies. First, when a property takes its values from a limited controlled set, the intermediate resource is materialized as an OWL named individual that is reused across all subjects sharing the same value. This is particularly useful for metadata such as instrument model, ionization polarity, or sample type, but it also applies to analytical descriptors such as library spectrum quality. Second, when values are numerical or drawn from an open set, intermediate resources are merged whenever they share both the same value and the same class (i.e., the same \texttt{rdf:type}). These deduplication strategies are deliberately not applied to PROV-O typed nodes (entities and activities), so that each analytical workflow remains a distinct and fully traceable provenance trail.

The PROV-O ontology ensures a clear separation of responsibilities within our knowledge graph by organising entities according to their PROV-O class (entity, activity, agent) and their type (metadata or data)~\cite{PROVOPROVOntology}, so the graph reflects the real-world processes that take place in a laboratory. The Mass Spectrometry ontology (MS) is used as the primary domain ontology~\cite{martensMzMLCommunityStandard2011}; although MS was originally developed for proteomics, we reuse and extend it for metabolomics-specific needs (e.g., the Feature concept). Additional ontologies are used for targeted metadata modeling: ChEBI for chemical entities, NCBITaxon for organism annotation, ENVO for environmental context, and NCIT for specific descriptors such as sample type and life stage~\cite{hastingsChEBIReferenceDatabase2013,federhenNCBITaxonomyDatabase2012,buttigiegEnvironmentOntologyContextualising2013,golbeckNationalCancerInstitutes2003}. This restricted but structured ontology set improves interoperability while preserving semantic coherence across the graph.

\begin{figure}
\centering
\includegraphics[width=0.95\linewidth]{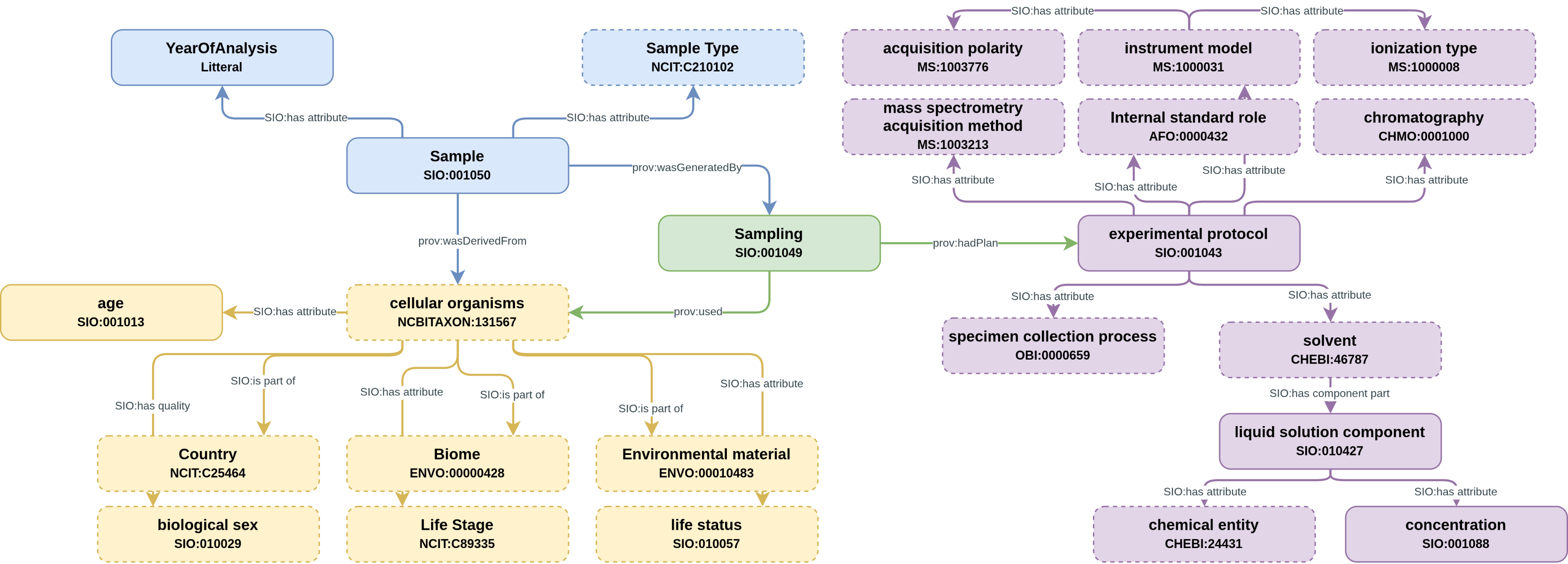}
\caption{\label{fig:metadata}Sample metadata layout linking sample description (blue), sampling process (green), source organism and environment (yellow), and experimental protocol (purple). In this and following figures: plain-border filled nodes are data entities and dashed-border nodes are individuals shown via their parent concept.}
\end{figure}

For the metadata design, we initially drew on the ISA-Tab model~\cite{johnsonISAAPIOpen2021}. However, several limitations emerged. Although an ISA-Tab ontology exists, its alignment with ontologies such as PROV-O, SIO, and BFO is provided through tabular mapping files rather than native semantic integration~\cite{johnsonISAAPIOpen2021,PROVOPROVOntology,dumontierSemanticscienceIntegratedOntology2014}. In addition, multiple \texttt{owl:Restriction} statements make partial reuse difficult, effectively requiring adoption of the full ontology when only a subset is needed. Our goal was to retain only the classes  \texttt{isa:Assay} and \texttt{isa:Study} as core entities.
In practice, the \texttt{isa:Assay} definition was too broad and oriented more toward biological objectives than our chemical focus. We also lacked sufficient metadata to distinguish assay types across experiments. For \texttt{isa:Study}, available metadata were likewise insufficient to describe sample processing, with less than 20\% coverage for collection and extraction fields~\cite{elabieadEnablingPanrepositoryReanalysis2025}. We therefore adopted 
a model focusing on sample description, source organism, and experimental protocol (Figure~\ref{fig:metadata}).

\begin{table}
\centering
\tiny
\caption{Ontology prefixes and namespace URIs used in MetaboKG.}
\label{tab:ontology_prefixes}
\begin{tabular}{ll}
\toprule
\textbf{Prefix} & \textbf{Namespace URI} \\
\midrule
prov & \url{http://www.w3.org/ns/prov#} \\
SIO & \url{http://semanticscience.org/resource/} \\
MS & \url{http://purl.obolibrary.org/obo/MS_} \\
ChEBI & \url{http://purl.obolibrary.org/obo/CHEBI_} \\
NCBITaxon & \url{http://purl.obolibrary.org/obo/NCBITaxon_} \\
ENVO & \url{http://purl.obolibrary.org/obo/ENVO_} \\
NCIT & \url{http://purl.obolibrary.org/obo/NCIT_} \\
dcat & \url{http://www.w3.org/ns/dcat#} \\
MBS & \url{https://ns.inria.fr/metaboKG/schema/} \\
\bottomrule
\end{tabular}
\end{table}

These choices fit the scope of MetaboKG, focused on chemical analysis rather than the full biological objective. Because collection and extraction metadata are too sparse for full process descriptions, they are represented as individuals attached to the experimental protocol; a dedicated handcrafted sampling process preserves the process-level connection in the workflow.

\begin{figure}
\centering
\includegraphics[width=0.99\linewidth]{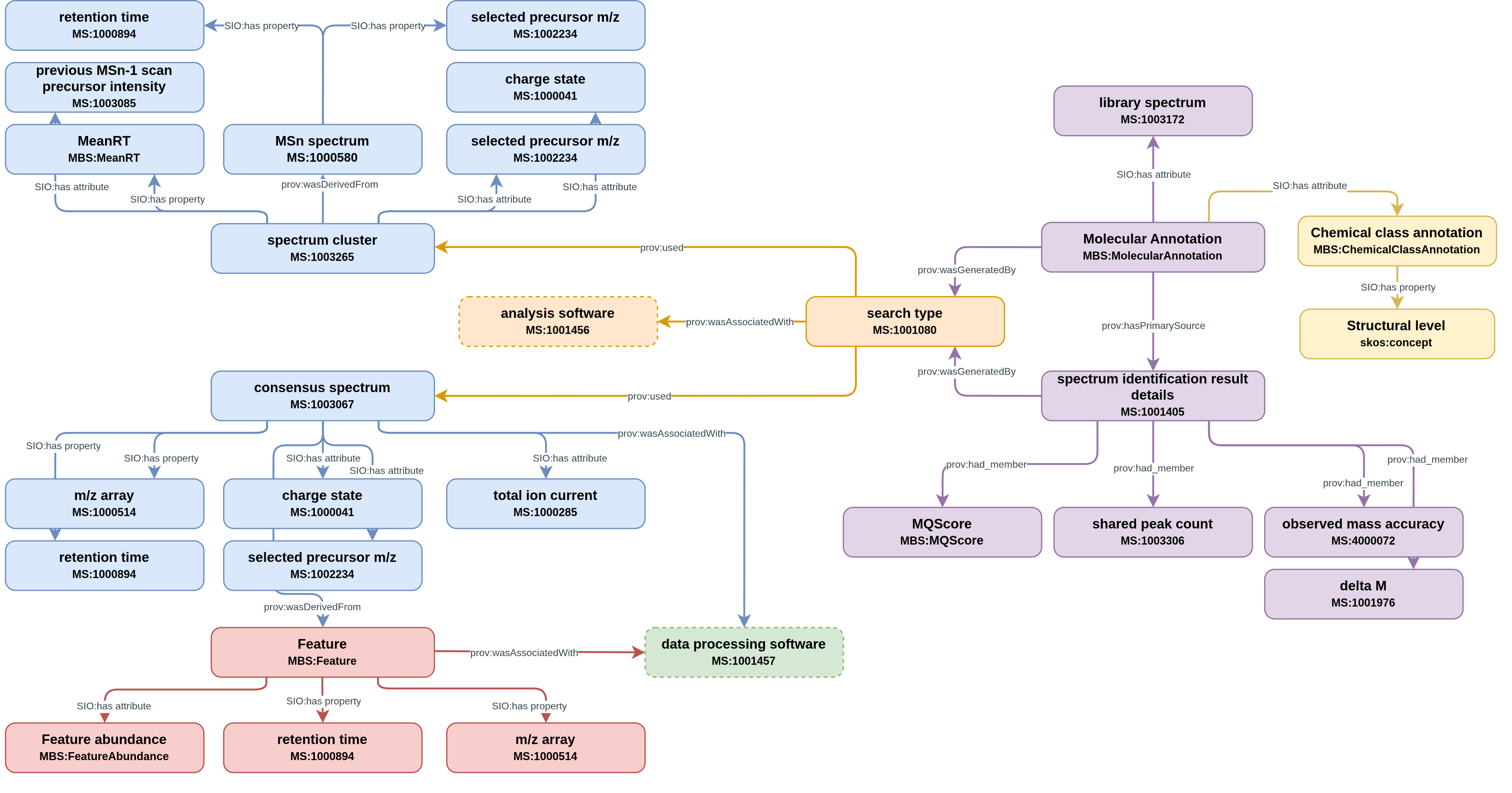}
\caption{\label{fig:Full_GNPS}Unified layout for GNPS Molecular Networking (MN, top) and Feature-Based Molecular Networking (FBMN, bottom): spectral data (blue), FBMN-specific features (red), annotation metadata (purple), annotation data including chemical class and structural level (yellow), and processing software (green).}
\end{figure}

\begin{figure}
\centering
\includegraphics[width=0.99\linewidth]{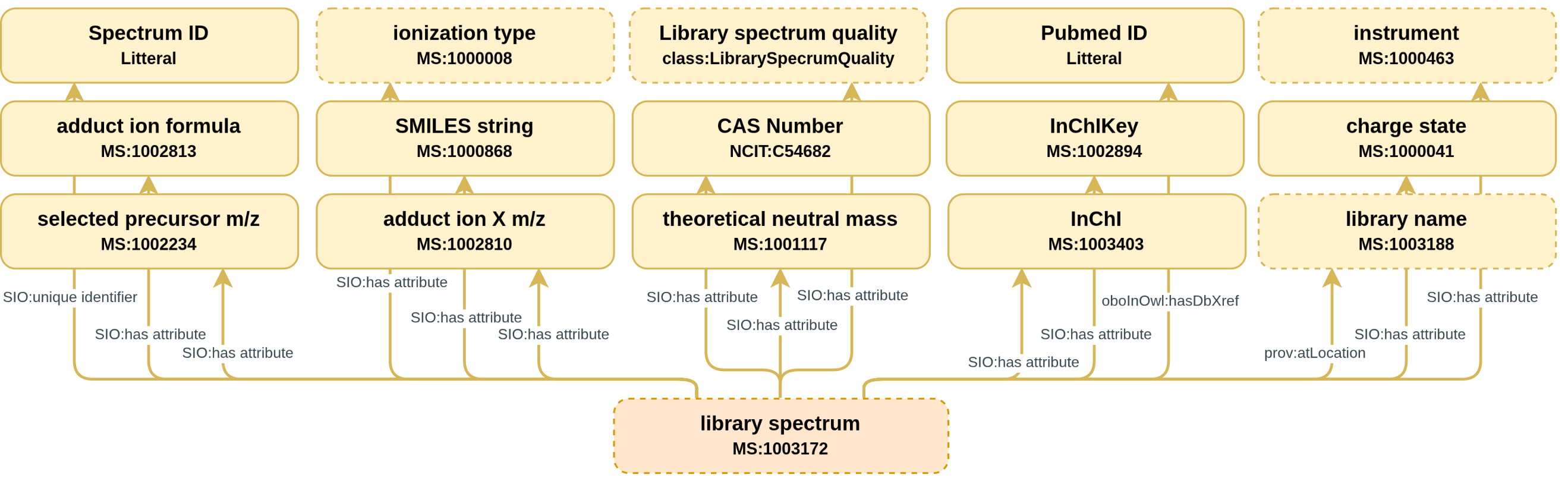}
\caption{\label{fig:LibrarySpec}Layout for a public spectral library spectrum, linking precursor and adduct information, molecular descriptors, acquisition and instrument context, resource metadata, and quality assessment.}
\end{figure}

The experimental part of the graph was designed to be compatible with current annotation methods, accurate yet usable, and source-preserving. Figure~\ref{fig:Full_GNPS} shows the unified layout for molecular networking (MN) and feature-based molecular networking (FBMN): annotation data (spectral and feature entities) are separated from annotation metadata (identification results and chemical classification), distinguishing empirical measurements from library-derived context and supporting provenance and FAIR principles. The public spectral library is modeled as unique spectrum individuals generated from GNPS results rather than imported from the source library, keeping only the metadata required for our study while staying synchronized with current GNPS outputs (Figure~\ref{fig:LibrarySpec}).

\section{Extracting Heterogeneous Metabolomics Resources into the MetaboKG Knowledge Graph}
\label{sec:collecting}

\subsection{Standardizing and Semantically Aligning Sample and Study Metadata}

\begin{figure}
\centering
\includegraphics[width=0.85\linewidth]{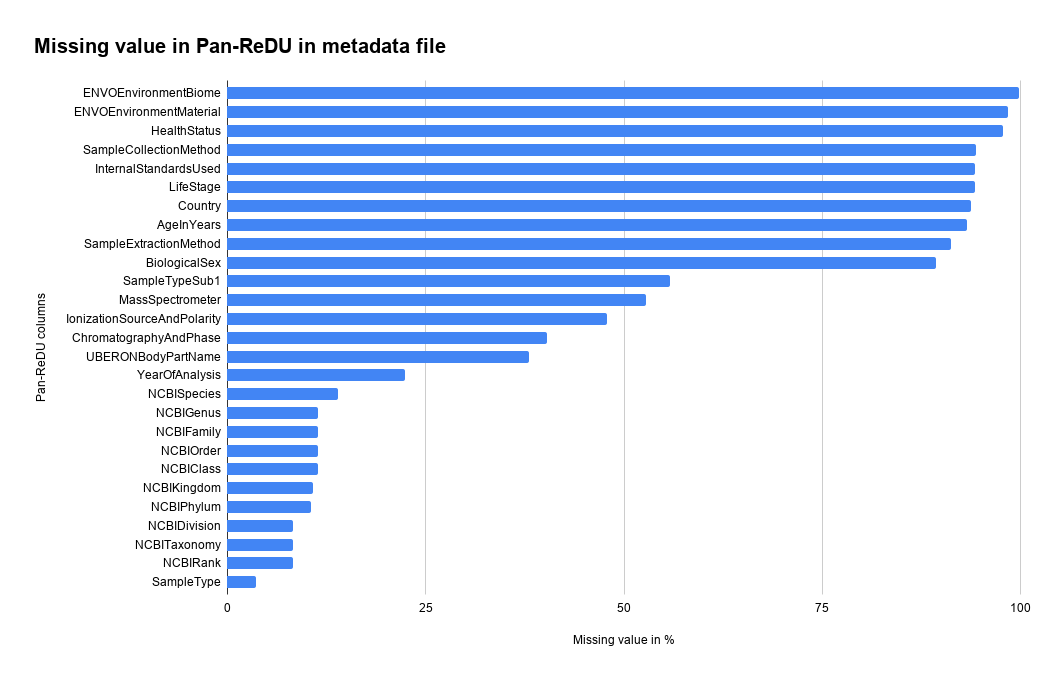}
\caption{Percentage of missing values per Pan-ReDU metadata field, sorted by missingness. Highly populated fields (e.g., SampleType, NCBI Taxonomy ranks) contrast with sparse ones (e.g., ENVOEnvironmentBiome, HealthStatus, LifeStage).} \label{fig:missing_metadata}
\end{figure}

To leverage standardized metadata, we utilized the Pan-ReDU resource, which harmonizes metadata into controlled vocabularies across major metabolomics repositories, including MetaboLights (MTBLS), the Metabolomics Data Repository (NMDR), and GNPS/MassIVE. This hub is refreshed regularly (\url{https://redu.gnps2.org/}) to provide a comprehensive cross-platform overview, and the resulting data matrix exhibits significant sparsity and uneven population. This disparity is primarily driven by the differing submission protocols of the source repositories; while MTBLS and NMDR enforce mandatory metadata reporting, GNPS allows for optional submissions. Consequently, the final Pan-ReDU metadata file contains a high degree of heterogeneity, with missingness rates exceeding 90\% for several fields (Figure~\ref{fig:missing_metadata}).

While many Pan-ReDU columns already use controlled vocabularies, we systematically mapped remaining uncontrolled terms to standardized ontologies using the Pan-ReDU Metadata Validation Sheet as a reference for defining individuals, applying three strategies: ``Reused'' for values kept as literals or direct ontology terms, ``Mapped'' for named individuals created via mapping dictionaries, and ``-'' for fields not integrated. Columns combining procedural and compositional information in a single cell (collection method, extraction method, internal standards) required custom parsing and curation before ontology alignment.

We used a semi-automated semantic alignment workflow on the EBI Ontology Lookup Service (OLS4)~\cite{mclaughlinOLS4NewOntology2025}, representing metadata values as named individuals rather than data properties to limit graph complexity while preserving semantic expressivity. For columns requiring ontology-based standardization (InternalStandardsUsed, IonizationSourceAndPolarity, SampleCollectionMethod, SampleExtractionMethod), searches were filtered to CHEBI, MS, NCIT, CHMO, and OBI; top-1 matches were too inaccurate due to lexical ambiguity, so we retrieved top-5 candidates and manually curated the most appropriate ontology class.

\begin{table}
\centering
\caption{Accuracy metrics for automated ontology term matching in the MetaboKG knowledge graph. This table evaluates the performance of automated semantic annotation for four ReDU metadata columns by comparing algorithmically predicted top-K ontology mappings against manually curated reference mappings. ``Macro'' denotes accuracy calculated independently for each metadata column then averaged, while ``Micro'' denotes accuracy calculated across all unique terms from all columns combined.}\label{tab:performance_results}
\resizebox{0.8\textwidth}{!}{%
\begin{tabular}{llcccc}
\hline
\textbf{Top-K} & \textbf{Averaging} & \textbf{SampleExtraction} & \textbf{SampleCollection} & \textbf{IonizationSource} & \textbf{Internal} \\
 &  & \textbf{Method} & \textbf{Method} & \textbf{And Polarity} & \textbf{Standards} \\
\hline
Top-1 & Macro & 90.70\% & 64.71\% & 50.00\% & 70.37\% \\
Top-5 & Macro & 90.70\% & 76.47\% & 50.00\% & 70.37\% \\
\hline
Top-1 & Micro & 77.78\% & 58.33\% & 50.00\% & 62.50\% \\
Top-5 & Micro & 80.00\% & 69.44\% & 50.00\% & 62.50\% \\
\hline
\end{tabular}
}
\end{table}

For compositional metadata, regex-based parsers decompose values into structured parameters. For example, ``methanol-water (4:1) + 0.1\% formic acid'' is parsed into three chemical components mapped to their CHEBI identifiers (methanol, water, formic acid), their proportions, and the acidic additive; similar strategies apply to other method-related columns.

Organism entities are composite individuals aggregating eight contextual metadata columns (NCBITaxonomy as organism type, Country, ENVOEnvironmentBiomeIndex, ENVOEnvironmentMaterialIndex, BiologicalSex, LifeStage, HealthStatus, AgeInYears). Each unique attribute combination defines one organism individual, deduplicating the tabular representation by 98.4\% while preserving biological and environmental context. Validated named individuals are then instantiated through RDF Mapping Language (RML) specifications, with each mapping yielding a JSON dictionary linking original metadata labels to their assigned URIs.

\subsection{Collecting and Structuring GNPS Analytical Outputs}
\label{subsec:gnps-outputs}
\begin{figure}
\centering
\includegraphics[width=0.75\linewidth]{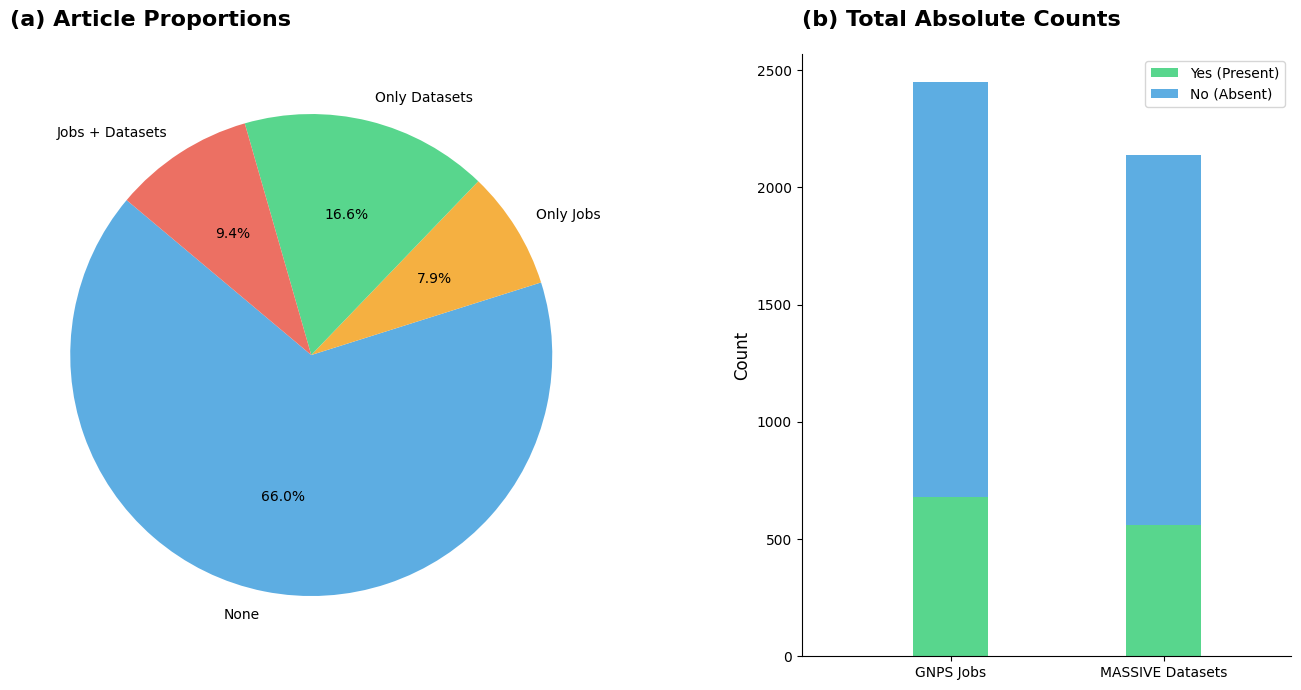}
\caption{\label{fig:GNPS_jobs_metrics}GNPS jobs and MASSIVE datasets in the analyzed PubMed literature. (a) Proportion of articles by data availability: none, GNPS only, MASSIVE only, or both. (b) Absolute frequency of presence (green, bottom) vs.\ absence (red, top) for GNPS jobs and MASSIVE datasets.}
\end{figure}

Most GNPS/MassIVE projects contain mostly unannotated raw datasets, with less than 5\% annotated, so direct ingestion of all public raw files is currently impractical for analysis-centric graph construction. We therefore start from annotated outputs and mine publicly shared GNPS job results identified through the literature. We processed 2,141 open-access papers citing FBMN or GNPS, converting them to Markdown using the Docling library and applying optimized regex patterns to extract workflow links and MassIVE IDs (Figure~\ref{fig:GNPS_jobs_metrics}). To ensure robust provenance and reproducible mapping, we bypassed the limitations of the GNPS API---which often lacks complete metadata paths---by downloading full job result bundles directly.

To link Pan-ReDU-style sample metadata with the corresponding GNPS analytical outputs, we use a large language model (LLM) to scan each paper and recover the explicit correspondence between MassIVE/dataset identifiers and the associated GNPS job identifiers. This step compensates for the fact that this mapping is rarely declared in the repository-side metadata and would otherwise have to be reconstructed manually. In a future version of the knowledge graph, raw data from the original collections---in particular MS/MS spectra---will be ingested alongside the annotated outputs to support spectrum-level reuse and re-annotation.

This process yields 680 GNPS jobs covering both molecular network (MN) and feature-based molecular network (FBMN) analyses, which produce CSV files with nearly identical column structure but distinct underlying logic. We harmonize them column by column using three mapping strategies: \emph{Direct map} (typed node carrying the table value as a \texttt{prov:value} literal), \emph{Literal} (untyped literal), and \emph{Ontology lookup} (reuse of an existing NamedIndividual when a match exists, otherwise creation under the listed ontology class).

\section{Improving Provenance Traceability and FAIR Accessibility}
\label{sec:provenance}

\begin{figure}
\centering
\includegraphics[width=0.75\linewidth]{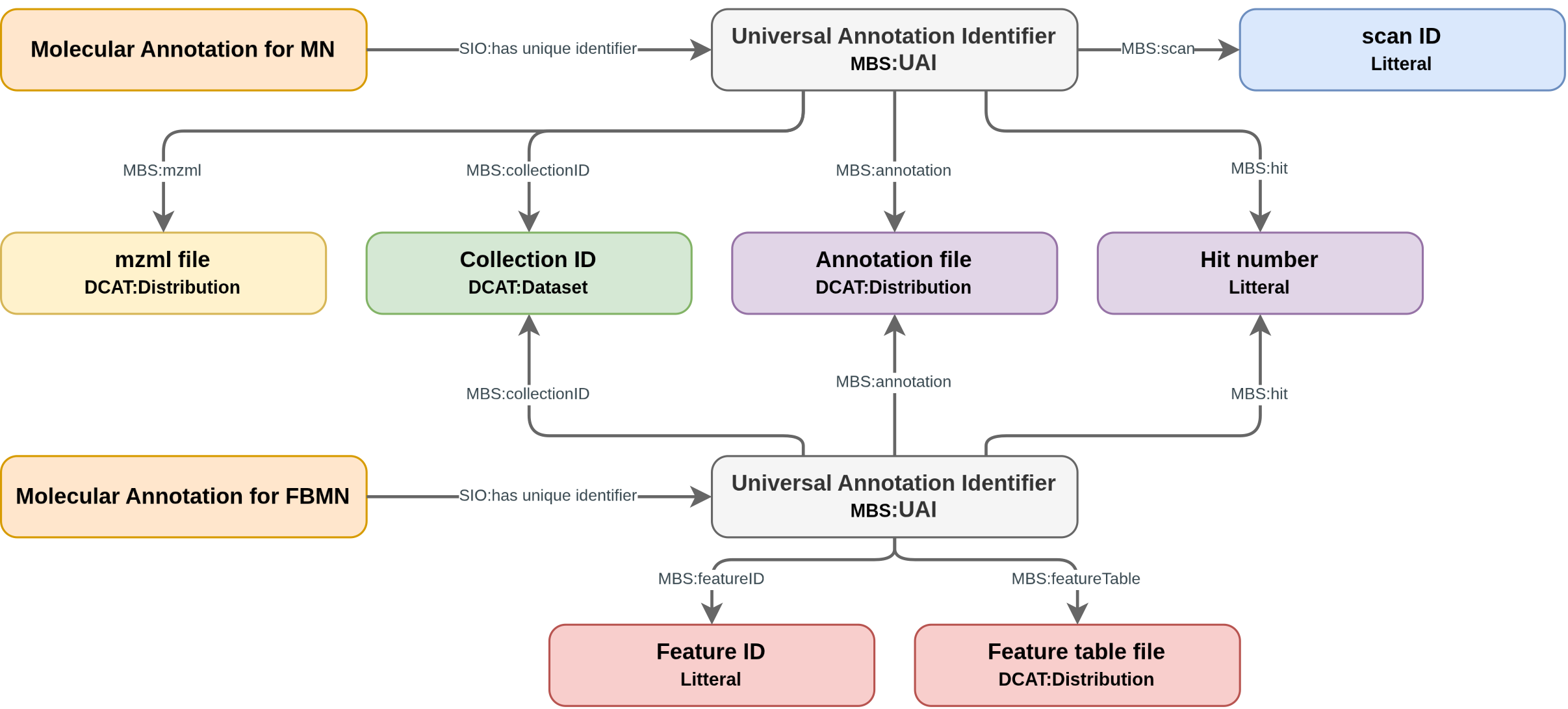}
\caption{Universal Annotation Identifier (UAI) schema. Each annotation links to a UAI node aggregating identifier components (collectionID, mzML, annotation file, hit number, featureID, feature table). In current GNPS workflows the hit number is always 1, since a single annotation is returned per cluster/spectrum; the field is kept for future ingestion of tools such as SIRIUS that report multiple candidates per spectrum. Files are represented as dcat:Distribution. Shared UAI components link MN and FBMN results post hoc, supporting incremental ingestion and FAIR access.} \label{fig:UAI}
\end{figure}

A central challenge in our workflow is integrating heterogeneous data sources distributed across multiple locations, while compound annotation is often repeated with different preprocessing pipelines and identification methods. To preserve provenance across these parallel analytical paths, we reused the Universal Spectrum Identifier (USI) and extended it with software-specific outputs and internal identifiers, enabling traceability of each identification back to its original dataset. The strategy decomposes identifiers into reusable components (collection, run/file, scan, hit, and feature level), providing consistent provenance capture for both MN and FBMN outputs and improving data accessibility (Figure~\ref{fig:UAI}).

First, we use DCAT to preserve the provenance of original files~\cite{rucknagelMetadataSchemaDescription2015}. Through \texttt{dcat:Distribution}, the model supports both local and remote files and can capture the public repository associated with each resource (omitted from Figure~\ref{fig:UAI} for readability).
Second, we introduce a new class, Universal Annotation Identifier (UAI), which centralizes identifiers in a single structure. Each PROV-O entity is linked to its own UAI. This identifier map addresses two key issues. (i) It enables post hoc linkage of outputs produced at different processing stages: entities can be connected by matching shared identifier components. For example, one result may include collectionID + mzML, while another includes collectionID + mzML + annotation; from the shared collectionID, we infer that both derive from the same original collection. (ii) It supports incremental data ingestion: datasets can be loaded first, and annotations can be added later while remaining linkable through identifier overlap. Because identifier fragments can be stored independently and completed over time (rather than requiring a single monolithic literal such as a fully populated USI), the model remains robust when metadata is sparse or partially missing.
Third, this design strengthens FAIR compliance, especially Findability and Accessibility. UAI provides a simple, stable, and queryable identifier layer, while dedicated properties improve targeted retrieval from any point in the graph. As a result, information is easier to discover and access without traversing complex graph paths manually.
Finally, UAI is designed for human-readable and persistent sharing across researchers. A UAI can be exchanged directly, avoiding the need to share long query templates or explain complex retrieval logic. This is particularly important for end users of the knowledge graph who are not familiar with advanced query authoring.

\subsection{URI Construction Strategy}

MetaboKG uses deterministic URIs following the pattern \textit{GlobalPrefix\/human-readable concept/\{hash\}}, balancing readability, stability, and robustness. The concept segment aids human understanding, while the hash ensures compactness. This departs from recommendations for fully opaque identifiers \cite{Sauermann2008CoolURIs,thalhammer2024cool}, but limits semantic risk by restricting interpretable parts to a stable modeling layer.

URIs reflect fine-grained scientific class, improving interpretability for analysis without losing compatibility. The hash uses \texttt{shortuuid}~\cite{korokithakisShortuuid} to keep URIs short and deterministic, especially when metadata are sparse. Hashes derive from distinguishing attributes where available, or from filename and source ID otherwise---ensuring stable, reproducible URIs without an external registry.

Random identifiers were avoided for reproducibility. This approach is a pragmatic middle ground: not globally governed, but compact, deterministic, and interpretable---suitable for reproducible mapping pipelines prioritizing provenance traceability and FAIR principles.

The resulting graph is stored and queried with OpenLink Virtuoso. We initially relied on Apache Jena Fuseki with a TDB backend, but query latency became prohibitive as the graph grew: with an average of about 10~million triples per ingestion batch, analytical SPARQL queries did not scale within acceptable response times. Migrating to Virtuoso provided substantially better performance on the same workload while keeping a standard SPARQL 1.1 endpoint, which is important since the competency-question evaluation relies on non-trivial SPARQL queries.

\section{Evaluation: Supporting Biochemical Exploration and Inference}
\label{sec:evaluation}

By integrating harmonized metadata, chemical annotations, instrument information, and provenance-aware workflow descriptions in a single semantic framework, MetaboKG supports cross-dataset reasoning over biochemical relationships that are difficult to detect when repositories are analyzed in isolation. We evaluate this through competency questions targeting biochemical enrichment, environmental specificity, life-stage associations, and cross-instrument analytical variation.

\paragraph{CQ1. Do GNPS annotations land on samples whose biological and environmental context has been harmonized in Pan-ReDU?}
Cross-repository integration is the central claim of MetaboKG: a molecular annotation produced by a GNPS workflow must be reachable from the corresponding sample, source organism, and acquisition protocol curated in Pan-ReDU. Answering this question requires resolving, for each annotation, the matching sample through the Universal Annotation Identifier and the shared collection identifier, and then traversing to the NCBITaxon source. This is what motivates the UAI design and the deliberate separation between analytical entities and contextual metadata in our semantic model.

\begin{lstlisting}[style=sparqlStyle, caption={SPARQL query implementing CQ1.}, label={lst:cq1}]
SELECT ?title (COUNT(DISTINCT ?ann) AS ?nAnn) (COUNT(DISTINCT ?samp) AS ?nSamp)
       (SAMPLE(?taxon) AS ?taxonExample)
WHERE {
  ?ann  a MBS:MolecularAnnotation ; SIO:000675 ?uA . ?uA MBS:collectionID ?coll .
  ?samp a SIO:001050 ;                  SIO:000675 ?uS . ?uS MBS:collectionID ?coll .
  ?coll dct:title ?title .
  OPTIONAL {
    ?samp prov:wasDerivedFrom ?src . ?src a ?taxon .
    FILTER STRSTARTS(STR(?taxon), "http://purl.obolibrary.org/obo/NCBITaxon_")
  }
}
GROUP BY ?title
ORDER BY DESC(?nAnn)
\end{lstlisting}

\paragraph{CQ2. How does spectral match quality vary across studies and across instruments?}
The reusability of an annotation pulled from a public repository depends on the modified cosine score, the number of shared peaks, and the mass accuracy of the underlying library match. End users need to be able to filter or weight annotations by their confidence at scale, not file by file. This requires that identification descriptors (MQScore, SharedPeaks, MZErrorPPM, MassDiff) be represented as first-class entities attached to the identification activity rather than as flat columns of a CSV, so that quality stratification can be expressed in a single SPARQL query across all ingested collections.

\begin{lstlisting}[style=sparqlStyle, caption={SPARQL query implementing CQ2.}, label={lst:cq2}]
SELECT ?title
       (AVG(xsd:decimal(?mq))  AS ?avgMQ)
       (AVG(xsd:decimal(?sp))  AS ?avgSharedPeaks)
       (COUNT(?ann) AS ?n)
WHERE {
  ?ann a MBS:MolecularAnnotation ;
       SIO:000675 ?u ; prov:hasPrimarySource ?ir .
  ?ir a MS:1001405 ;
      prov:had_member ?mqNode , ?spNode .
  ?mqNode a MBS:MQScore ; prov:value ?mq .
  ?spNode a MS:1003306    ; prov:value ?sp .
  ?u MBS:collectionID ?coll . ?coll dct:title ?title .
}
GROUP BY ?title
ORDER BY DESC(?avgMQ)
\end{lstlisting}

\paragraph{CQ3. For a given annotation, are the ClassyFire and NPClassifier taxonomies consistent, and which pairs co-occur most often?}
ClassyFire and NPClassifier provide complementary views of an annotated compound: structural chemistry on one side, biosynthetic pathway on the other. When stored side-by-side in repository-native tables, comparing the two taxonomies requires bespoke joins per file. In the graph, each annotation is linked to its full taxonomic profile, which makes the cross-classification queryable and aggregable across all collections at once. This question motivates the choice of typing chemical class annotations through SIO and reusing both taxonomies as parallel branches off the same molecular annotation node.

\begin{lstlisting}[style=sparqlStyle, caption={SPARQL query implementing CQ3.}, label={lst:cq3}]
SELECT ?cfClassLabel ?npcPathwayLabel (COUNT(DISTINCT ?ann) AS ?n)
WHERE {
  ?ann a MBS:MolecularAnnotation ;
       SIO:000008 ?cca , ?ncn .
  ?cca SIO:000223 MBS:CF_Class ; prov:value ?cfClassLabel .
  ?ncn SIO:000223 NPC:Pathway       ; prov:value ?npcPathwayLabel .
}
GROUP BY ?cfClassLabel ?npcPathwayLabel
ORDER BY DESC(?n)
\end{lstlisting}

\paragraph{CQ4. In which sample types has a given compound been observed across the integrated repositories?}
A central question of reverse metabolomics~\cite{gentryReverseMetabolomicsDiscovery2024,charron-lamoureuxGuideReverseMetabolomics2025} is, given an InChIKey, to retrieve every public study where the molecule has been annotated, together with the sample type and source organism in which it was detected. Answering this requires joining annotation-side spectral identifiers (library InChIKey) with sample-side descriptors (ReDU sample type and subtype, NCBITaxon, ENVO biome) through the same UAI bridge, and counting distinct collections rather than distinct files. This is precisely the kind of cross-dataset exploration that the analysis-centric design of MetaboKG is intended to support, and it would be impractical without a unified semantic layer over GNPS and Pan-ReDU.

\begin{lstlisting}[style=sparqlStyle, caption={SPARQL query implementing CQ4.}, label={lst:cq4}]
SELECT ?sampleType ?ik
       (COUNT(DISTINCT ?samp)  AS ?nSamples)
       (COUNT(DISTINCT ?title) AS ?nStudies)
WHERE {
  ?ann a MBS:MolecularAnnotation ; SIO:000675 ?uA ; SIO:000008 ?lib .
  ?lib a MS:1003172 ; SIO:000008 ?ikn .
  ?ikn a MS:1002894 ; prov:value ?ik .
  ?uA  MBS:collectionID ?coll . ?coll dct:title ?title .

  ?samp a SIO:001050 ; SIO:000675 ?uS ; SIO:000008 ?sampleType .
  ?uS   MBS:collectionID ?coll .
  FILTER STRSTARTS(STR(?sampleType),
                   "https://ns.inria.fr/metaboKG/schema/sampletype_")
  FILTER (!STRSTARTS(STR(?sampleType),
                     "https://ns.inria.fr/metaboKG/schema/sampletype_sub_"))
}
GROUP BY ?sampleType ?ik
HAVING (COUNT(DISTINCT ?title) > 3)
ORDER BY DESC(?nSamples)
LIMIT 500
\end{lstlisting}

In existing workflows, addressing such competency questions would require ad hoc scripts and file transformations across heterogeneous repositories, which are rarely FAIR-compliant. MetaboKG provides a FAIR-compliant alternative: the same questions are answered within a single SPARQL-queryable semantic framework, independently of the original file formats.

\section{Conclusion}
\label{sec:conclusion}

Untargeted metabolomics relies on large MS/MS volumes distributed across multiple public repositories, yet their scientific value depends on integrating analytical outputs with reusable metadata while preserving provenance, annotation confidence, and experimental context. Existing repository practices and tabular exchange formats only partially address this need, rarely preserving both analytical detail and metadata structure in a unified, interoperable representation.

We presented MetaboKG, an analysis-centric knowledge graph framework that transforms heterogeneous metabolomics resources into a common semantic structure through three coordinated elements: a transformation workflow for repository outputs and metadata; a semantic model centered on provenance, analytical entities, and contextual metadata; and an identifier strategy supporting traceability and late binding across workflows and data sources. Together they integrate GNPS and MassIVE data while preserving the information needed for analysis, and they show that current semantic web languages and ontologies---suitably combined---can cover the integration needs of semi-structured MS/MS data, answering competency questions that require joint access to chemical annotations, biological context, environmental descriptors, and workflow provenance.

A key perspective for future work is to adopt \texttt{LinkML}~\cite{moxonLinkMLOpenData2026} as a single source of truth for the MetaboKG schema. The current pipeline maintains four parallel artifacts---an OWL ontology, named-individual schemas, hand-written RML templates, and Python URI-minting code---that must stay coherent under schema evolution and are prone to drift each time a new annotation tool or column is added. Generating OWL, RML, Pydantic dataclasses, JSON Schema, and SHACL shapes from a single LinkML specification would make this drift mechanically impossible, replace ad hoc RML concatenation with declarative composition, and align MetaboKG with life-science schemas such as NMDC and MIxS, opening a path toward schema-level federation.

\subsubsection*{Source Code Availability.} The MetaboKG source code is available at \url{https://github.com/HolobiomicsLab/MetaBoKG}.

\subsubsection*{Acknowledgements.} This work was supported by the French government through the France 2030 investment plan via the MetaboLinkAI bilateral project (ANR-24-CE93-0012-01), the 3IA C\^ote d'Azur programme (ANR-23-IACL-0001), and ANR-22-CPJ2-0048-01. This work received financial support from the CNRS through the MITI interdisciplinary programs. Additional support was provided through the UCAJEDI Investments in the Future project (ANR-15-IDEX-01).

\subsubsection*{Author Contributions.} M.F., F.G., and L.-F.N.\ conceptualised the method. M.F.\ implemented the method and software and performed the analysis. F.G.\ and L.-F.N.\ co-supervised the project. M.F., F.G., and L.-F.N.\ wrote and edited the manuscript.

\subsubsection*{Use of AI-Assisted Technologies.} During the preparation of this work, the authors used large language models (including tools from Anthropic and OpenAI) to accelerate software development and to improve the readability and style of the manuscript. No LLMs were used to generate scientific content, results, or analyses beyond their described role for specific tasks (dataset curation) and their use is documented in the code. All AI-generated suggestions for code and text were critically reviewed, verified, and edited by the authors, who take full responsibility for the content of the article.

\subsubsection*{Disclosure of Interests.} The authors have no competing interests to declare that are relevant to the content of this article.

\bibliographystyle{splncs04}
\bibliography{MetaboKG}

\end{document}